\documentclass{aa}
\usepackage{graphics}

\newcommand{\kms}{\mathrm{\,km\,s}^{{-1}}}
\newcommand{\Teff}{T_\mathrm{eff}}
\newcommand{\FeH}{\mathrm{[Fe/H]}}
\newcommand{\masy}{\mathrm{\,mas\,y}^{-1}}
\newcommand{\Mv}{M_\mathrm{V}}
\newcommand{\logg}{\log g}
\newcommand{\Blim}{B_\mathrm{lim}}
\newcommand{\Ha}{H_\mathrm{\alpha}}
\newcommand{\vr}{v_\mathrm{r}}
\newcommand{\pc}{\,\mathrm{pc}}
\newcommand{\kpc}{\,\mathrm{kpc}}
\newcommand{\K}{\,\mathrm{K}}
\newcommand{\vsini}{v\sin i}

\begin{document}
\thesaurus{10               
		(10.15.2 NGC 2355;   
individual:
                 08:11.1;   
		 08.18.1;   
		 08.06.3 )} 

\title{Fundamental properties of the open cluster NGC 2355}

\thanks{based on observations made on the 193cm telescope at 
the Haute-Provence Observatory, France, and on plate digitisation at the Centre
d'Analyse des Images, Paris.
This publication makes use of data products from the Two Micron All Sky
Survey, which is a
joint project of the University of Massachusetts and the Infrared
Processing and Analysis
Center, funded by the National Aeronautics and Space Administration and the
National Science
Foundation.}

\author{C. Soubiran, M. Odenkirchen, J.-F. Le Campion }

\offprints{C. Soubiran}

\institute{Observatoire de Bordeaux, BP 89, F-33270 Floirac, France}

\date{Received ; accepted }

\titlerunning{NGC 2355}
\maketitle

\begin{abstract}
NGC 2355 is an old open cluster in the outer part of the galactic disk
($l=203\fdg4$,
$b=+11\fdg8$) which has been little studied until now.
This paper presents the first astrometric and spectroscopic investigation
of this cluster. We have measured precise absolute proper motions from old
Carte du Ciel plates, POSS-I plates and recent CCD observations obtained
with the Bordeaux meridian circle. The proper motion data reveal
38 highly probable cluster members down to $\Blim = 15$~mag within $7'$
of the cluster center.
We have also obtained ELODIE high resolution spectra for 24 stars.
Seventeen of them are confirmed to be members of the
cluster on the basis of radial velocity. Eight of them are fast rotating
turnoff stars for which the projected rotational velocity has been determined.
The spectroscopic observations have also provided estimates of the physical
parameters $\Teff, \logg, \FeH, \Mv $ of the 24 target stars. Two
stragglers have been
identified in the cluster. Combining our astrometric and spectroscopic
results with previous UBV photometry and recent JHK$_s$ photometry from
the 2MASS survey we have derived the fundamental properties of the cluster:
metallicity, age, distance, size, spatial velocity and orbit.

\keywords{Galaxy : open clusters and associations: individual: NGC 2355 --
               stars : kinematics --
               stars: rotation --
               stars: fundamental parameters}

\end{abstract}

\section{Introduction}

Old open clusters, with ages greater than the age of the Hyades ($\sim$ 600 Myr),
represent a minority of about 80 objects among 1200 known open clusters.
Among their properties which enable to investigate both stellar physics and
galactic structure
(reviewed by Friel 1995), we are especially interested in orbits because
they are related to the
processes which have allowed them to survive tidal forces.
The statistics are still poor but it seems that old open clusters follow
orbits that keep them away
from the plane and the disruptive effects of giant molecular clouds.
The question is to know if these orbits result from special events or
represent the tail of the
distribution of clusters that have already been destroyed. Another relevant
point to clarify is the
relationship between orbits and metallicity [Fe/H] which traces the
dynamical and chemical
evolution of the Galaxy.
The metallicity of old open clusters is intermediate between the disk and
the thick disk,
with a radial gradient, but a large dispersion that could
indicate an inhomogeneous enrichment of the Galaxy. To answer such fundamental 
questions, new observations are needed to investigate in more details the old
open cluster properties and their correlations. We have 
therefore undertaken
a spectroscopic and astrometric program to obtain metallicities, distances
and velocities
of high quality for several poorly known old open clusters, NGC 2355 being
our first target.\

There are very few references on NGC 2355 in the astronomical literature. A
photometric study
in UBV down to $V\sim 19.2$ was made by Kaluzny \& Mazur (1991). In this
study,
the reddening of the cluster was estimated to be $E_{B-V}$=0.12 mag, the
distance modulus $(m-M)_0=12.1$, the metallicity +0.13  and
the age the same as Praesepe. In their search for old open clusters, Phelps et
al. (1994) also report a
photometric study of NGC 2355 in BV but the photometry of individual
stars is not given. Their
calibration of the index $\delta V$, defined as the magnitude difference
between the
main-sequence turnoff and the giant clump leads to a Morphological Age
Index corresponding to 0.9~Gyr,
like Praesepe (Janes \& Phelps 1994). More recently, Ann et al. (1999) examined this cluster
as part of the BOAO survey (Bohyunsan Optical Astronomy Observatory, Korea) and 
determined from UBVI photometry : $\FeH=-0.32, E_{B-V}=0.25, (m-M)_0=11.4$ and an 
age of 1 Gyr.\

In Sect. 2 and 3 we present new data which are used to analyse the cluster
in combination with the
UBV photometry of Kaluzny \& Mazur (1991) and the JHK$_s$ photometry which is
available for the whole field in the 2MASS 1999 Spring Incremental Data
Release.
We describe the determination and analysis of proper motions from photographic
plates and recent observations at the meridian circle of Bordeaux (Sect.
2). For
24 bright stars ($V \le 13$) around the cluster's center, spectra were
obtained
with the echelle spectrograph ELODIE on the 193cm telescope at the
Haute-Provence Observatory.
The radial velocities of the red giants were obtained by standard on-line
reduction directly at the
telescope.  The determination of the radial velocity and projected
rotational velocity of the hot
fast rotating turnoff stars required dedicated reduction
tools (Sect. 3).
In Sect. 4 we present our analysis of the spectra to estimate the
atmospheric parameters
$\Teff, \logg, \FeH $ and the absolute magnitude $\Mv$. For the latter, we
developed a new version of the TGMET method (Katz et al. 1998 and Soubiran et al.
1998).
We discuss the case of an unusual giant in NGC 2355 which is 2.3 magnitudes
brighter than the giant clump
for the same temperature. We also report the discovery of a blue straggler
in the cluster and of a
moving pair of field stars. Sect. 5 deals with the fundamental parameters
of NGC 2355 resulting from our
study. Our conclusions are reviewed in Sect. 6. For identifying individual
stars we use
as far as available the star numbers introduced by Kaluzny \& Mazur (1991),
preceded by the prefix "KM".

\section{Measurement and analysis of proper motions}

\begin{figure*}[t]
\resizebox{18cm}{!}{\includegraphics{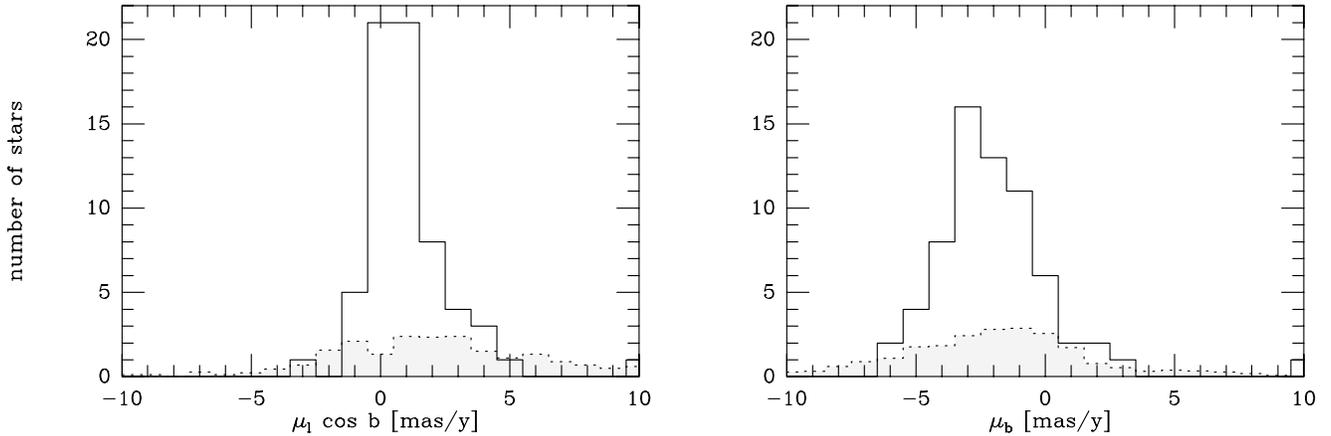}}
\caption{Distribution of stellar proper motions (in galactic components
$\mu_l,\mu_b$) in a circular field of $7'$ radius around the center of the
cluster (solid line) and in an annulus with radii $18'$ and $36'$
(dotted line). The number counts in the annulus are rescaled to the surface
of the circle in order to enable direct comparison.}
\label{}
\end{figure*}

We determined precise proper motions for stars in the cluster region
and in the surrounding field for two reasons: 1. to enable
a kinematical segregation of cluster members and non-members, and 2. to
derive the absolute tangential velocity of the cluster.
In order to achieve adequate accuracy, the proper motions were determined from
observations at 3 epochs with a maximum separation of about 90 years.
The first epoch, around 1910, was provided by a triple-exposed plate from the
Bordeaux Carte du Ciel (CdC, $2^\circ \times 2^\circ$, $\Blim \simeq 15.0$) on
which the cluster is favorably placed near the plate center, and by
5 plates from the Bordeaux Astrographic Catalogue (AC, same size,
$\Blim \simeq 12.0$) which fully or partially overlap with the
field of the CdC-plate. Second epoch positions were obtained by
measurement of two POSS-I glass copies (O \& E plates) from the Leiden Observatory plate
archive.
The third epoch consists of observations made with the CCD meridian circle
of Bordeaux Observatory in 1997/1998 as part of the `M\'eridien 2000' program
(see Colin et al. 1998).

The CdC and POSS plates were scanned on the MAMA
machine at the Paris Observatory, the AC plates were scanned on a PDS machine at the
Astronomical Institute M\"unster. The scans were processed partly with the
SExtractor software (Bertin \& Arnouts 1996) and partly with our own
software, in
particular for the centering of the CdC triple images. All observations were
reduced to the reference system of Hipparcos. The meridian observations and
the first-epoch plate measurements were linked directly to reference stars
from the Hipparcos catalogue and the ACT Reference catalogue (Urban et al. 1998). 
Iterative reduction schemes were used
in order to make due account of the multiple observations of each star.
The measurements from the POSS-I plates required a separate treatment because
the geometry of projection on these plates is subject to complicated
and sizable distortions. Thus we constructed from the first and third-epoch
data an intermediary catalog of secondary reference stars.
We then applied a moving-filter technique as described by Morrison
et al. (1998) to transform the POSS-I data locally and smoothly to the
Hipparcos system.  The final proper motions were obtained by combining
the positions from all epochs in a weighted linear least-squares adjustment.

According to the reduction residuals and the comparison between different
observations of the same epoch, the mean accuracies of the positions are
as follows: 50 mas per coordinate for the mean positions from the meridian
circle observations, between 120 and 150 mas per coordinate and plate for the
positions from the first-epoch plates and 150 mas per coordinate and plate
for the positions from POSS-I. The mean internal errors of the resulting
proper motions range from $0.7 \masy$  for the brightest stars to $2.0 \masy$
for the faintest stars of the sample.

Without selection according to kinematics, the distribution of the stars in
the
plane of the sky reveals that the cluster is centered on the position
$\alpha = 7^\mathrm{h}17\fm0, \delta = 13^\circ45'$ (2000.0), and that its
angular radius is at least $5'$, but probably larger.
In Fig.~1 we present histograms of the distribution of proper motions for a
circular field of $7'$ radius around the above given position.
For comparison we also show the distribution of proper motions in an
annulus outside
the cluster, namely between $18'$ and $36'$ from the center (counts
rescaled to equal surface).
It is clearly seen that the cluster stands out against the field as a
concentration of
comoving stars.

In order to estimate the mean proper motion of the cluster and to obtain
cluster membership probabilities we fitted two-dimensional Gaussians to the
proper-motion distributions of the pure field sample and the
cluster-and-field sample. The parameters of the distributions were
determined by
applying a maximum-likelihood criterion to the proper motions in the
range $\mu_l\cos b\in [-6,+10]\masy$ and $\mu_b\in [-10,+6]\masy$.
The distribution of the
field stars appears centered around $(\mu_l\cos b,\mu_b) = (+2.4,-1.8) \masy$
and has a dispersion of about $3\masy$. The proper motions of the cluster
stars
are centered on $(\mu_l\cos b,\mu_b) = (+0.5,-2.4)\masy$, i.e. they are
slightly
offset from the mean proper motion of the field, and have
a dispersion of $0.8\masy$ in $\mu_l\cos b$ and $1.5\masy$ in $\mu_b$.
The dispersion in $\mu_b$ is in agreement with the estimated mean proper
motion errors. However it is surprising that the dispersion in $\mu_l\cos
b$ is
substantially smaller. The determination of the mean proper motion of the
cluster
has a statistical uncertainty of $0.3\masy$. To this we must add in
quadrature the uncertainty of the absolute calibration of the Hipparcos
reference frame
which is $0.25\masy$ (Kovalevsky et al. 1997). With some additional
allowance for
other (possibly undetected) systematic errors in the measuring process
we estimate that the accuracy of our determination of the absolute proper
motion of the
cluster is $0.5\masy$ per coordinate.

Using the above given parameters for the distributions of the proper
motions in the cluster
and the field, individual kinematical membership probabilities were
calculated.
This was done in the usual way according to the relative frequency of
cluster stars which the
fitted model distributions predict for a given proper motion.
Our sample of stars in the $7'$ circle thus divides into 38 probable
cluster members ($p > 90\%$),
13 probable field stars ($p < 10\%$) and 17 unclear cases ($10\% \le p \le
90\%$).

By kinematical discrimination between cluster stars and field
stars one obtains an improved picture of the structure and spatial extent
of the cluster.
For this purpose we chose a field of $36'$ radius around the cluster
center and selected only those stars with proper motion equal to the mean
motion of the
cluster within $1\sigma$, i.e. the proper motion dispersion of the cluster
stars.
The latter criterion reduces the surface density of the field stars by a
factor of 10,
but retains a sufficiently large number of cluster stars so that the
cluster's structure and extent become more clearly recognizable.
Fig.~2 compares the radial profile of stellar density (number counts in
non-overlapping annuli around the cluster center) with and without
kinematical selection. It turns out that the cluster has a central
component with exponentially decreasing density out to about $7'$, a halo with
approximately constant density beyond $7'$ and an edge at $15'$.
The core radius of the cluster, i.e. the radius at which the surface
density drops to half its central value, is found to be about $1.5'$.

\begin{figure}[h]
\resizebox{8.5cm}{!}{\includegraphics{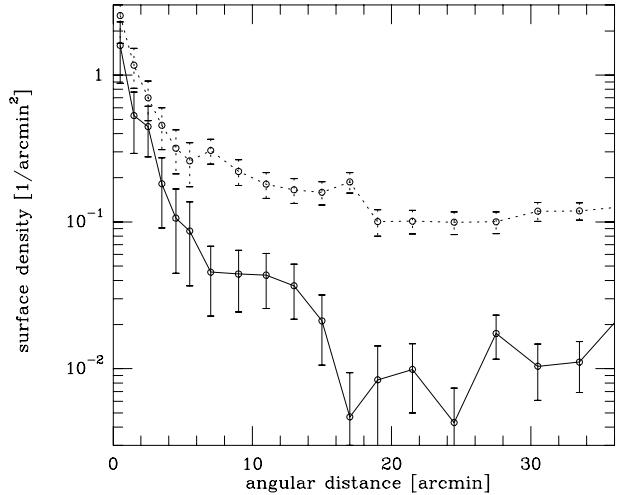}}
\caption{Radial density profile of the cluster as obtained from star counts
down to
$B = 15.0$ in non-overlapping annuli around the cluster center. Dotted
line: without
kinematic selection. Solid line: only stars with proper motion close to the
mean motion
of the cluster. Error bars show the size of $\sqrt{N}$ variations.}
\label{}
\end{figure}

\section{Spectroscopy : radial and rotational velocities}

Twenty-four stars with $V \le 13$ in the field of the cluster were observed at
the 
Haute-Provence Observatory in January 1999, on the 1.93m telescope equipped with the
spectrograph ELODIE.
This instrument is a dual-fibre-fed echelle spectrograph devoted to the
measurements of accurate
radial velocities (Baranne et al. 1996).  A spectral range 390-680 nm is
recorded in a single
exposure as 67 orders on a 1K CCD at a mean resolving power
of 42000.
With a one hour exposure one typically achieves a S/N of 100 on a star of
magnitude 8.5 or a S/N of
10 on a star of magnitude 12.8.  ELODIE is a very stable instrument,
allowing to compare easily spectra
observed at different epochs.

Optimal extraction and wavelength calibration are automatically performed
on-line, as well as the
measurement of radial velocities by digital cross-correlation with binary
templates
thanks to the TACOS reduction software developed by D. Queloz (1996). The
cross-correlation technique
is well adapted for strong-lined spectra, corresponding to moderate
effective temperatures up to 6500 K.
The precision of radial velocities for such spectral types is better than
$100\,\mbox{m s}^{-1}$ even at
low signal to noise ratio. Among the 24 target stars, 16 stars presented a
clean deep correlation profile,
permitting an accurate radial velocity and FWHM measurement directly at
the telescope.
The observations revealed a strong concentration of stars at $\vr \sim
35\kms$.
This is without doubt the trace of the radial motion of the cluster.
Thus, on the basis of radial velocity, 9 target stars were confirmed to be
members of the cluster, with colours
corresponding to clump giants. The mean radial velocity of this sub-sample
is $35.13\kms$ with a standard
deviation of $0.39\kms$. The FWHM of the correlation function was for most
of the stars nearly constant
at $11.3\kms$, but slightly larger for 3 stars (KM 1 : $13.2\kms$ , KM 2
: $34.6\kms$,
KM 20 : $17.3\kms$) corresponding to the signature of either
macroturbulence, rotation or binarity.
The radial velocities are listed in Tab. 1, together with the UBV
photometry from Kaluzny \& Mazur (1991)
and the JHK$_s$ photometry from 2MASS.

Eight stars with bluer colour presented broad lines indicating a high
rotational velocity. They could not
be treated by the standard cross-correlation method. Instead, their rotational
profile was extracted using a
least-squares deconvolution technique developed and fully described by 
Donati et al. (1997). 
The latter method presents some similarities with the cross-correlation
method. It is based on the
fact that the observed spectrum can be expressed as the convolution product
of a line pattern
with a rotational plus instrumental profile. This profile can thus be
recovered by deconvolving the observed
spectrum with a line mask computed from a model atmosphere having the same
parameters as the star.
As the effective temperatures of the target stars were not known, a series
of line masks with $\Teff$ ranging
from 6000 K to 9000 K were computed from the Kurucz's database (Kurucz 1993).
The best contrast was obtained at $\Teff \sim 7000 - 7500\K$. The
deconvolution
was quite difficult due to the low signal to noise ratio of the spectra but
the signature of the rotation
was visible for each star and confirmed a radial velocity consistent with
the cluster's velocity. The
next step was to calibrate the width of the deconvolved
profiles in terms of $\vsini$. For this task we used several reference stars
for which both a high-quality ELODIE spectrum and a published value of
$v\sin i$ were available.
Nine  stars from the TGMET library (see next section) were found in the
catalogue of rotational velocities
compiled by Uesugi \& Fukuda (1982), restricted to $50 - 200\kms$.
The FWHM of the deconvolved profiles were measured
the same way for reference and target stars by fitting a 10 degree
polynomial as can be seen in Fig. 3. In
this example, the same line mask corresponding to the parameters $\Teff =
7500 K, \logg = 4.0, \FeH = 0.0$ was
used for the deconvolution of the two spectra, but HD~132052 ($\vsini =
120\kms$) was observed at S/N=131 while
KM~13 was observed at S/N=11. The weighted linear regression performed on
the 9 points given by the reference
stars led to the relation $\vsini = 0.443 \cdot \mbox{FWHM} + 19.5 \kms$,
represented in Fig. 4. The projected rotational
velocities estimated for the 8 turnoff stars of NGC 2355 are given in
column 10 of Tab. 1.

 \begin{figure}[h]
\resizebox{8.5cm}{!}{\includegraphics{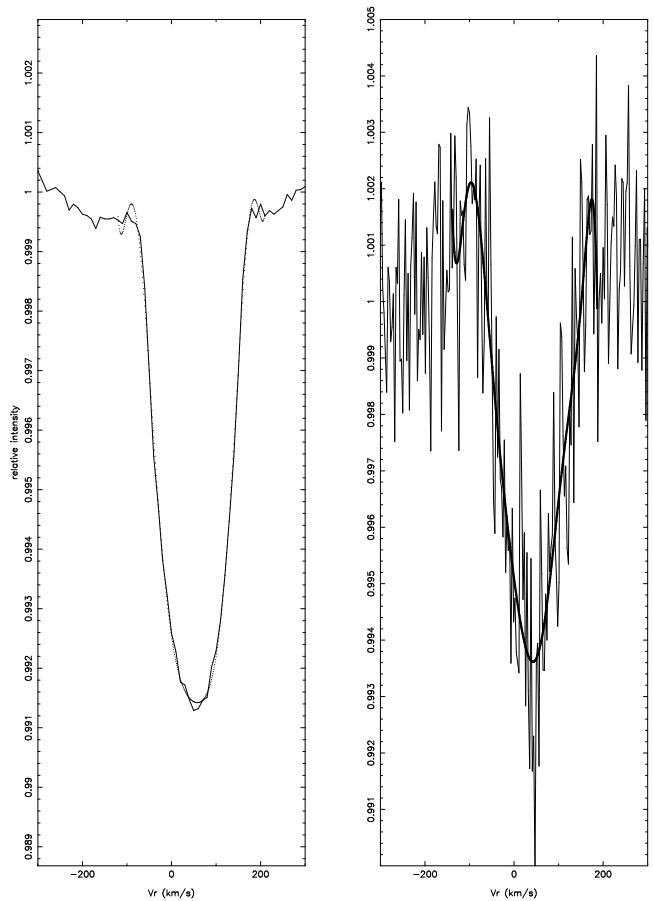}}
\caption{The deconvolved profiles of HD 132052 ($\mbox{S/N} = 131, \vsini =
120\kms$) and KM 13 (S/N=11)
are fitted with a 10 degree polynomial.  The fit on KM 13 gives $\vr =
40\kms$ (heliocentric),
$\vsini = 90\kms$}
\label{}
\end{figure}

\begin{figure}[h]
\resizebox{8.5cm}{!}{\includegraphics{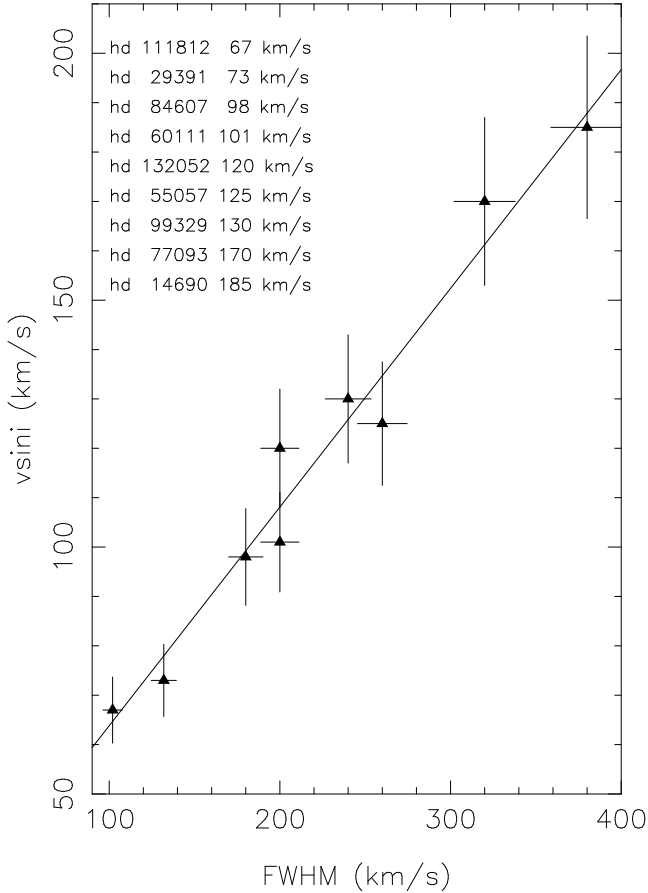}}
\caption{A linear relation between the measured FWHM of the deconvolved
profiles and the projected
rotational velocities was computed from 9 reference stars observed also
with ELODIE, with known projected
rotational velocities from Uesugi \& Fukuda (1982). An error of 20\% was
assumed on the projected rotational
velocities taken from the litterature, and the horizontal error bars
represent the standard error on the
measurement of the FWHM. }
\label{}
\end{figure}

The most remarkable star among those which are classified as cluster
members in Tab. 1 is KM 1.
This is a giant which is 2.3 magnitudes brighter than the giant clump of
the cluster.
The fact that it is within one sigma of the cluster distribution in
position, proper motion
and radial velocity makes us believe that it is a cluster member and not a
field star.
On the other hand, we cannot completely exclude the possibility that it is
a projection of a field star
onto the cluster because a field star at this position may by chance have
the same radial velocity
as the cluster. According to the formulae of the circular motion around the 
galactic center, a star on the line of sight to the cluster at a distance
between $0.5\kpc$ and $2\kpc$
would have a heliocentric radial velocity
between $+15\kms$ and $+27\kms$.
Hence there is a certain overlap between the radial velocity distribution
of the field stars and the
radial velocity of the cluster.

If KM 1 is a member of the cluster, then it remains unclear by which
phenomenon this star is considerably
brighter than the other cluster giants of the same colour. We have looked
for photometric variations to
check if this star could be in an unstable phase of its evolution.
No variations could be detected in the meridian observations over 3 years,
nor in comparison with the apparent brightness on the 1950 POSS-I plates
and the 1910 CdC plate.
KM 1 is part of the TYCHO catalogue (TYC 775 997 1) where no variability is
reported. The spectrum
of this peculiar star is discussed in more detail in the next section.

Another remark is to be made on the star KM 2.
This one was selected by Ahumada \& Lapasset (1995) as a blue straggler
candidate.
However, its radial velocity indicates that this star is not a member of
the cluster.

We also point towards the stars KM~20 and KM~26. These are found to be
field stars with
identical radial velocities of $\vr = 50.3\kms$, despite an angular
separation of
$1.7'$. The hypothesis of a moving pair is discussed at the end of the next
section.

\begin{table*}[h]
\caption[]{List of the 24 stars in the field of NGC 2355 observed with ELODIE.
UBV photometry is from Kaluzny \& Mazur (1991)(*: CCD meridian V
photometry), JHK$_s$ is from
2MASS.  S/N is the mean signal to noise ratio of the spectrum at 550 nm.
$\vr$ (heliocentric) results from the standard on-line reduction and  $\vsini$ 
from the least-square deconvolution method (see text).}
\begin{flushleft}
\begin{tabular}{| l | r | r | r | r | r | r | r | r | r | c |} \hline
object & V & B$-$V & U$-$B & J & H$-$K$_s$ & J$-$K$_s$ & S/N  & $\vr$  &
$\vsini$ & member \\ \hline
KM  1  &  9.847 & 1.059 & 0.834 &  7.952  &  0.135 &   0.576 & 16 & 35.57 &
- & yes\\
KM  2  & 11.896 & 0.367 & 0.102 & 11.077 &   0.012 &   0.177 & 12 & 10.22 &
- & no \\
KM  3  & 12.152 & 0.950 & 0.828 &  9.845 &   0.115 &   0.494 & 16 & 34.79 &
- & yes\\
KM  4  & 12.247 & 1.004 & 0.729 & 10.360 &   0.172 &   0.561 & 14 & 34.89 &
- & yes\\
KM  6  & 12.612 & 1.146 & 1.050 & 10.537 &   0.129 &   0.673 & 9.5& 61.22 &
- & no \\
KM  7  & 12.619 & 1.011 & 0.681 & 10.705 &   0.096 &   0.612 & 11 & 19.98 &
- & no \\
KM  8  & 12.663 & 1.027 & 0.762 & 10.738 &   0.142 &   0.597 &  9 & 35.73 &
- & yes \\
KM  9  & 12.692 & 1.025 & 0.759 & 10.769 &   0.060 &   0.550 & 14 & 35.06 &
- & yes \\
KM 10  & 12.736 & 1.176 & 0.965 & 10.504 &   0.167 &   0.665 & 19 & 35.03 &
- & yes \\
KM 11  & 12.755 & 1.064 & 0.749 & 10.734 &   0.152 &   0.622 & 14 &
$-$32.43 & - & no \\
KM 12  & 12.760 & 0.396 & 0.242 & 11.844 &   0.100 &   0.177 & 5.5& - & 90
& yes  \\
KM 13  & 12.809 & 0.341 & 0.210 & 12.033 &   0.033 &   0.201 & 11 & - &  90
& yes  \\
KM 14  & 12.829 & 0.330 & 0.193 & 12.032 &   0.034 &   0.144 &  7 & - &
120 & yes  \\
KM 15  & 12.852 & 1.128 & 1.095 & 10.838 &   0.115 &   0.662 & 17 & 34.72 &
- & yes \\
KM 19  & 12.958 & 0.412 & 0.255 & 12.013 &   0.071 &   0.197 & 18 & - & 120
& yes   \\
KM 20  & 12.999 & 1.164 & 0.841 & 10.800 &   0.138 &   0.605 & 13 & 50.29 &
- & no  \\
KM 21  & 13.091 & 0.432 & 0.206 & 12.138 &   0.019 &   0.231 &  8 & - & 105
& yes  \\
KM 22  & 13.136 & 0.394 & 0.243 & 12.240 &$-$0.017 &   0.156 & 15 & - & 125
& yes  \\
KM 26  & 13.466 & 1.100 & 0.812 & 11.382 &   0.209 &   0.636 & 14 & 50.31 &
- & no \\
KM 27  & 13.550 & 0.394 & 0.273 & 12.713 &   0.010 &   0.147 & 13 & - & 105
& yes  \\
GSC 77500538 & 12.09* & - & - &10.280  &  0.149 &   0.597 &  7 &    35.68 &
- & yes \\
GSC 77501060 & 12.09* & - & - & 9.669   & 0.155 &   0.798 & 21 & $-$42.62 &
- & no  \\
GSC 77501198 & 12.57* & - & - &10.865   & 0.096 &   0.506 & 12 &    34.70 &
- & yes\\
GSC 77501264 & 11.76* & - & - &11.259  &  0.019 &   0.113 &  9 & - &  115 &
yes  \\ \hline
\end{tabular}
\end{flushleft}
\end{table*}

\section{Atmospheric parameters, absolute magnitudes}

The atmospheric parameters ($\Teff, \logg, \FeH $) and the absolute
magnitude $\Mv$ have been
obtained with the automated software TGMET, described in Katz et al.
(1998).   TGMET is a minimum
distance method (reduced $\chi^2$ minimisation) which measures in a quantitative
way the similarities and
discrepancies between spectra and finds for a given target spectrum the
most closely
matching template spectra in a library. The TGMET library (Soubiran et al. 1998) 
was built with high S/N ELODIE spectra of
reference stars for which the atmospheric parameters were taken from
published detailed analyses, mostly in the Catalogue of [Fe/H] 
determinations (Cayrel de Strobel et al. 1997). The previous version of the
TGMET library was extended to cover the temperature interval [3500 K - 7500 K]
and now includes nearly 450 reference spectra of all metallicities. To improve 
the temperature estimation,
the library was also completed with stars having reliable $\Teff$, either from
the list of Blackwell \& Lynas-Gray (1998)
based on ISO flux calibration, or from the calibration of the colour index
V$-$K (Alonso et al. 1996 and
Alonso et al. 1999).  A new aspect of TGMET was developed by estimating
the absolute magnitude $\Mv$
simultaneously with the atmospheric parameters, based on
the fact that stars having similar
spectra have similar absolute magnitudes. In fact most of the stars of TGMET
library are in the Hipparcos catalogue and 313 of them have parallaxes with a relative 
errors lower than 10\%. Stars from the library having precise Hipparcos parallaxes 
had their absolute magnitude $\Mv$ derived from the TYCHO V apparent magnitude. They were
used as reference stars for the absolute magnitude as for the atmospheric 
parameters. Some tests were performed to check the reliability of such
spectroscopically determined absolute magnitudes. At solar metallicity, the rms
difference between the absolute magnitude determined from Hipparcos and from TGMET is
0.21 for dwarfs, 0.31 for clump giants and 0.50 for other giants. The parameters 
($\Teff, \logg, \FeH , \Mv $) of a 
target star processed by TGMET are given by the weighted mean of the parameters
of the best matching reference spectra (presenting a reduced $\chi^2$ which does
not exceed the lowest one by more than 12\%). The resulting uncertainty depends 
mainly on two factors. The first one is the
quality of the parameters found in the literature for the reference stars. Typically
errors quoted in detailed analyses are  
50 to 150 K for
$\Teff$, 0.1 to 0.3 for $\logg$, and  0.05 to 0.1 for $\FeH$. But for some reference stars,
it happens that the errors on the atmospheric parameters from the literature are much higher 
and such stars 
are gradually being identified with TGMET as outliers. The spectral detailed 
analysis is the only primary 
method to estimate metallicities. It is a difficult task and the results obtained by 
different authors can differ by a large amount. We do not expect to do better than detailed 
analyses with TGMET, but
the method will improve if we can add reference stars with very reliable atmospheric
parameters to the library . The problem is not as critical for absolute magnitudes because 
the large majority of
the reference stars have excellent Hipparcos parallaxes thus reliable absolute magnitudes.
The second source of uncertainty in the TGMET results is the way the
parameter space is sampled by the reference stars. For example, among evolved stars,
results are expected to be better for clump giants than for other giants because clump
giants are more numerous in the literature and Hipparcos, consequently better
represented in the TGMET library than other kinds of giants, and also because clump
giants occupy a smaller volume than other giants in the parameter space.\

The results of TGMET for the 24 target stars are given in Tab. 2.
The last column presents the distance moduli $(V-\Mv)$ of the stars as
derived from
the spectroscopically determined $\Mv$.
To illustrate the TGMET processing Fig. 5 shows the spectrum of KM~10  in
the region of the MgI triplet together
with its best matching reference spectrum HD~205435  $(\Teff = 5068 K,
\logg = 2.64, \FeH = -0.16, \Mv = 1.097)$.
As another example, Fig. 6 represents the $\Ha$ line of the fast
rotator KM 22 together with its best
matching reference spectrum HD~201377 ($\Mv = 1.607$).

\begin{figure}[h]
\resizebox{8.5cm}{!}{\includegraphics{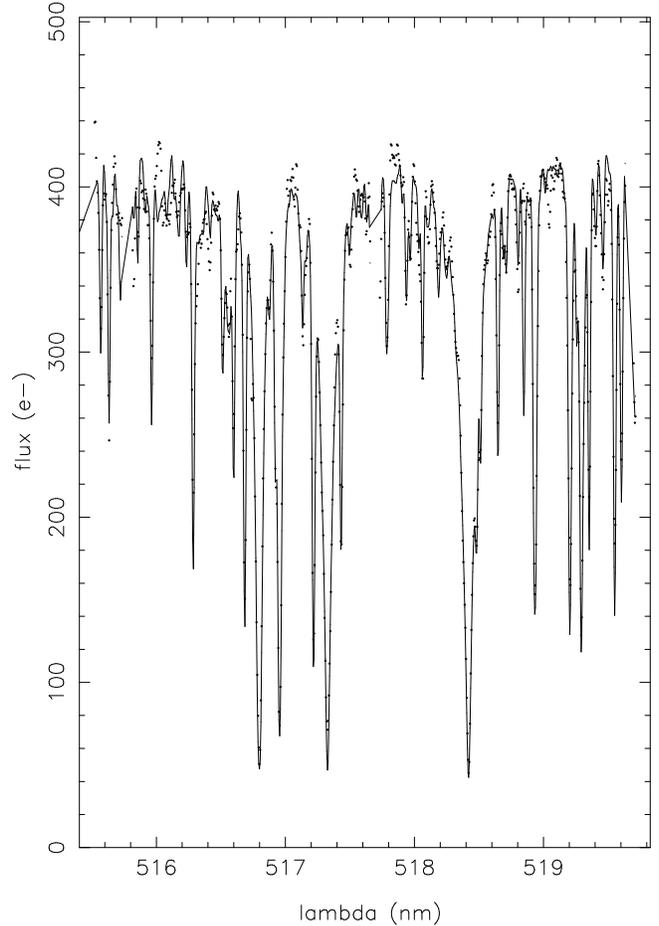}}
\caption{The spectrum of the clump giant KM~10 (dots) in the region of the MgI
triplet 
together with the best matching reference spectrum, HD~205435 (S/N=164).
The reference spectrum was
 put at the same  radial velocity scale and flux level as the target. }
\label{}
\end{figure}

\begin{figure}[h]
\resizebox{8.5cm}{!}{\includegraphics{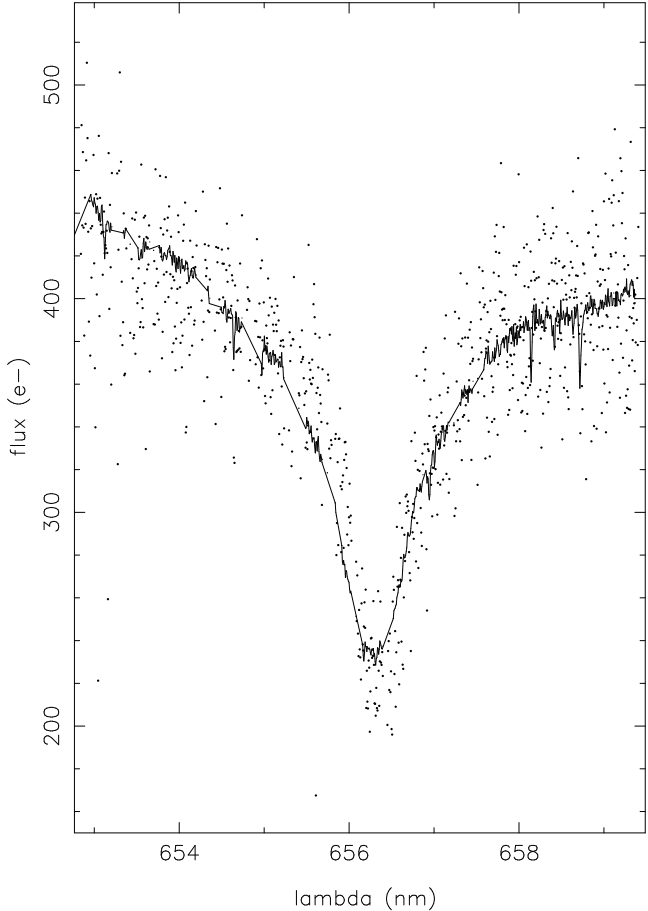}}
\caption{Same as Fig. 5 for the fast rotator KM~22 and  HD~201377 (S/N=167)
in the
region of  $\Ha$.}
\label{}
\end{figure}

In the previous section, it was pointed out that KM~1 is surprisingly found
to be a probable member of the cluster despite
a visual magnitude much brighter than the clump giants, for the same colour.
Its distance modulus is consistent with the rest of the cluster and hence
in agreement with the supposed membership,
as well as its metallicity. Nevertheless this star has an abnormal, thus
interesting, position on the cluster's HR
diagram which is worth attention (see Figs. 7 and 8).
The intrinsic physical difference between KM 1 and the other giants of the
cluster has been investigated by
comparing their respective TGMET best matching reference stars, the
parameters of which are presented in
Tab. 3. The two sets present similar mean temperatures and metallicities
but the range of absolute magnitudes
is quite different. KM 1 exhibits through TGMET more spectral
similarities with supergiants like HD~215665 or HD~159181 than with clump giants.
The larger dispersion which is found for the parameters of KM 1, especially for 
$\Mv$, is well explained by the fact that the TGMET library does not sample the 
parameter space at the same resolution for supergiants than for clump giants as
already mentioned. The weighted mean and error bar for each parameter of KM 1 
was computed with the 10 best-fitting reference stars, despite large differences,
because they equally (within 12\%) match the target spectrum. The large error bars 
reflect the fact that there is not a perfect analog of KM 1 in the TGMET library.
Consequently, in the following, KM 1 will contribute with a lower weight than clump giants
to the determination of the fundamental parameters of the cluster. 
The 20 reference stars listed in Tab. 3, except HD~214567, are reported in
Uesugi \& Fukuda (1982) to rotate
at $\vsini \sim 10-20\kms$, so that there is no difference between the two
sets concerning the rotation.
The brighter magnitude of KM~1 could correspond to a higher mass, but in
that case KM~1 should be much younger than
the other giants. By comparison to the isochrones of Girardi et al. (2000)
in the plane ($\Teff, \Mv$), KM~1 should be 160 Myr old whereas the rest of
the cluster is 1 Gyr old (see next section).
This phenomenon is similar to the blue staggler phenomenon, but on the red,
evolved side.
Ahumada \& Lepasset (1995) enumerate several theories which have been
proposed for the blue stagglers, and which could
also explain the observation of KM~1: a field star captured by the cluster,
a star which
has accreted mass from the interstellar medium, a star which formed after
the bulk of the cluster members,
the result of a non-standard mechanism in the evolution, the result of a
stellar collision or a binary merger.
KM~1 could also be a blue straggler which has evolved. At the present
state, the main difference observed
between the spectra of KM~1 and the clump giants is a slightly broadened 
profile as seen
on macroturbulent supergiants, or on rotating or binary 
giants, and a difference in absolute magnitude detected by TGMET.
A spectrum with much higher S/N is necessary in order to obtain further
insight on the nature of this star.

Also mentioned in the previous section, KM~20 and KM~26 might be a moving
pair because they have a common
radial velocity of $50.3\kms$. Their metallicities of $-0.26$ and $-0.31$
are in agreement.
Unfortunately, the other parameters determined for KM~20 present large
standard errors (see Tab. 2) indicating that the fit with the
TGMET reference spectra was not satisfactory. We recall that KM~20 has an
enlarged profile with
$\mathrm{FWHM}=17.3\kms$. Thus this star might be a spectroscopic binary
which would explain the mediocre
results of TGMET. Since a precise estimate of the proper motion of KM~26 is
missing (no measurement due to the blending
of the image by a reseau line of the CdC plate) the common motion cannot be
assessed by means of proper motions.
Anyway, the proper motion of KM~20 is small ($\mu_l\cos b = -2.2\masy,
\mu_b = -1.6\masy$), so a large distance
is probable and the verification of common proper motion would be difficult.
Based on the spectroscopic estimate of the distance of KM~26 (2.7 kpc), the
angular separation corresponds to a
linear distance between the two stars of $1.3\pc$.

\begin{table*}[h]
\caption[]{ Atmospheric parameters $(\Teff, \logg, \FeH)$ and absolute
magnitude $\Mv$ obtained
with TGMET, and standard errors. Column 7 indicates the
reference star which was found to have the most similar spectral
characteristics over the considered
wavelength interval. The last column gives the distance modulus $(V-\Mv)$. }
\begin{flushleft}
\begin{tabular}{| l | c | c | c | r | r | l | r |} \hline
object & member & $\Teff $ & $\logg $ & $\FeH $ & $\Mv$  & best reference &
$V-\Mv$ \\ \hline
  KM  1    & yes &$4988 \pm   70 $ & $ 2.16 \pm  0.19 $ & $ -0.04 \pm  0.04
$ & $ -1.611 \pm  0.417$& HD 203387
   &11.451 \\
  KM  2    & no  &$6835 \pm   54 $ & $ - $ & $ -0.28 \pm  0.04
$ & $  2.768 \pm  0.215$& HD  18995
   & 9.122 \\
  KM  3    & yes &$4987 \pm   36 $ & $ 2.72 \pm  0.07 $ & $ -0.14 \pm  0.06
$ & $  0.559 \pm  0.096$& HD  27022
   &11.591 \\
  KM  4    & yes &$4961 \pm   35 $ & $ 2.67 \pm  0.06 $ & $ -0.15 \pm  0.08
$ & $  0.572 \pm  0.079$& HD   5395
   &11.668 \\
  KM  6    & no  &$4835 \pm   60 $ & $ 2.84 \pm  0.12 $ & $ -0.17 \pm  0.04
$ & $  1.138 \pm  0.288$& HD 185351
   &11.472 \\
  KM  7    & no  &$4898 \pm   40 $ & $ 2.65 \pm  0.07 $ & $ -0.50 \pm  0.08
$ & $  0.998 \pm  0.222$& HD 127243
   &11.612 \\
  KM  8    & yes &$5122 \pm   78 $ & $ 2.75 \pm  0.10 $ & $ -0.23 \pm  0.06
$ & $  0.950 \pm  0.380$& HD 198809
   &11.710 \\
  KM  9    & yes &$4995 \pm   32 $ & $ 2.64 \pm  0.05 $ & $ -0.12 \pm  0.05
$ & $  0.392 \pm  0.197$& HD 198809
   &12.298 \\
  KM 10    & yes &$4995 \pm   31 $ & $ 2.76 \pm  0.08 $ & $ -0.15 \pm  0.05
$ & $  0.694 \pm  0.274$& HD 205435
   &12.036 \\
  KM 11    & no  &$4893 \pm   50 $ & $ 2.65 \pm  0.06 $ & $ -0.50 \pm  0.06
$ & $  1.193 \pm  0.199$& HD 219615
   &11.557 \\
  KM 12    & yes &$7254 \pm   92 $ & $ 4.30 \pm  0.20 $ & $  0.02 \pm  0.02
$ & $  1.769 \pm  0.261$& HD  99329
   &10.991 \\
  KM 13    & yes &$7187 \pm  131 $ & $ 4.30 \pm  0.20 $ & $ -0.08 \pm  0.10
$ & $  1.763 \pm  0.231$& HD 193581
   &11.037 \\
  KM 14    & yes &$7175 \pm  133 $ & $ 4.14 \pm  0.20 $ & $ -0.11 \pm  0.08
$ & $  1.813 \pm  0.240$& HD 169032
   &11.007 \\
  KM 15    & yes &$4687 \pm   82 $ & $ 2.77 \pm  0.11 $ & $  0.00 \pm  0.03
$ & $  1.353 \pm  0.350$& HD 222404
   &11.497 \\
  KM 19    & yes &$7386 \pm   45 $ & $ - $ & $  -
$ & $  1.394 \pm  0.291$& HD 169032
   &11.556 \\
  KM 20    & no  &$5126 \pm  120 $ & $ 3.02 \pm  0.20 $ & $ -0.26 \pm  0.09
$ & $  1.793 \pm  0.438$& HD  27022
   &11.197 \\
  KM 21    & yes &$7053 \pm  166 $ & $ - $ & $ -0.28 \pm  0.25
$ & $  1.836 \pm  0.239$& HD  99329
   &11.254 \\
  KM 22    & yes &$7385 \pm   46 $ & $ - $ & $  -
$ & $  1.387 \pm  0.289$& HD 169032
   &11.743 \\
  KM 26    & no  &$4908 \pm   41 $ & $ 2.68 \pm  0.08 $ & $ -0.31 \pm  0.06
$ & $  0.843 \pm  0.267$& HD 188119
   &12.617 \\
  KM 27    & yes &$7362 \pm   50 $ & $ - $ & $  0.06 \pm  0.06
$ & $  1.654 \pm  0.203$& HD 193581
   &11.896 \\
  GSC 500538& yes &$4953 \pm  36 $ & $ 2.66 \pm  0.06 $ & $ -0.22 \pm  0.07
$ & $  0.488 \pm  0.126$& HD  27022
   &11.602 \\
  GSC 501060& no  &$4359 \pm  62 $ & $ 2.02 \pm  0.13 $ & $ -0.17 \pm  0.04
$ & $ -0.143 \pm  0.472$& HD   5234
   &12.233 \\
  GSC 501198& yes &$5024 \pm  57 $ & $ 2.66 \pm  0.10 $ & $ -0.14 \pm  0.09
$ & $  0.432 \pm  0.083$& HD 198809
   &12.138 \\
  GSC 501264& yes &$7287 \pm 117 $ & $ 4.30 \pm  0.20 $ & $  0.03 \pm  0.03
$ & $  1.332 \pm  0.319$& HD 201377
   &10.428 \\
\hline
\end{tabular}
\end{flushleft}
\end{table*}

\begin{table}
\caption[]{ Mean parameters from the literature and Hipparcos of the
reference spectra matching the best the target spectra of  KM~1 and 
clump giants of NGC 2355.
The parameters of KM 1 were computed from the weighted mean 
of these 10 reference stars which equally (within 12\%) fit the
KM 1 spectrum. The 10 reference stars which are presented for cluster's 
clump
giants correspond to those which occur the most often in the TGMET 
solution of KM 3, KM 4, KM 8, KM 9, KM 10, KM 15, GSC 500538 and
GSC 501198. }
\begin{tabular}{|lccrr|}\hline
\noalign{\smallskip}
 name   &  $\Teff$   &  $\logg$   & $\FeH$   &  $\Mv$  \\
\hline
 \noalign{\smallskip}
 \multicolumn{5}{|c|}{----------  KM~1  ----------}  \\
 HD 203387& 4959   &   2.74   &   $-$0.05   &     0.183  \\
 HD 210807& 4949   &   2.52   &   $-$0.19   &   $-$0.527  \\
 HD 215665& 4835   &   2.47   &   $-$0.09   &   $-$1.470  \\
 HD 198809& 5087   &   2.88   &   $-$0.14   &     0.461  \\
 HD 185758& 5303   &   2.91   &   $-$0.15   &   $-$1.433  \\
 HD 159181& 5169   &   1.54   &      0.16   &   $-$2.425  \\
 HD 206859& 4604   &   1.56   &      0.00   &   $-$2.881  \\
 HD  26630& 5183   &   1.40   &   $-$0.03   &   $-$2.577  \\
 HD 209750& 5156   &   1.34   &      0.17   &   $-$3.889  \\
 HD   3712& 4611   &   2.04   &   $-$0.10   &   $-$2.003  \\
 \noalign{\smallskip}
 \hline
 \noalign{\smallskip}
 \multicolumn{5}{|c|}{------  clump giants  ------} \\
 HD 198809& 5087   &   2.88   &   $-$0.14   &    0.461  \\
 HD  27022& 5100   &   2.47   &      0.05   &    0.241  \\
 HD 214567& 4994   &   2.70   &      0.03   &    0.481  \\
 HD 188119& 4885   &   2.61   &   $-$0.32   &    0.592  \\
 HD 205435& 4950   &   2.64   &   $-$0.16   &    1.097  \\
 HD   5395& 4770   &   2.55   &   $-$0.70   &    0.628  \\
 HD 135722& 4810   &   2.56   &   $-$0.44   &    0.718  \\
 HD 210807& 4949   &   2.52   &   $-$0.19   & $-$0.527  \\
 HD 222404& 4778   &   2.98   &   $-$0.01   &    2.510  \\
 HD  35369& 4873   &   2.50   &   $-$0.26   &    0.489 \\
\noalign{\smallskip}
\hline
\end{tabular}
\end{table}

\section{Fundamental parameters of the cluster}
\subsection{Metallicity, age, reddening, distance, size}
The weighted average of [Fe/H] of the members of NGC 2355 listed in Tab. 2
gives a metallicity of
$\FeH = -0.07 \pm 0.11$.  The  standard error on $\FeH$ does not take 
into account the uncertainties on $\FeH$ for the reference stars which are 
usually unknown. For example, the reference clump giant HD 5395 (see Tab. 3), with 
$\FeH = -0.70$, has been 
used to derive the parameters of KM 3, KM 4 and GSC 500538, despite an uncertain metallicity :  
Fernandes-Villacanas et al. (1990) report $\FeH=-1.0$, McWilliam (1990) 
$\FeH=-0.51$ whereas TGMET gives a value of $\FeH=-0.44$. This illustrates the 
difficulty to define a precise reference system for the atmospheric parameters.
 In this
light, the error bar on [Fe/H] is underestimated.
We have estimated the age of NGC 2355 by chosing in the isochrones of solar
abundance of Girardi et al.
(2000) the one which matched the best our observations in the plane
($\Teff,\Mv$). Fig. 9 shows that an age of
1 Gyr is probable due to the position of the turnoff stars.

Despite an unexpectedly dispersed photometry, as previously mentioned by
Kaluzny \& Mazur (1991), Ann et al. (1999) and confirmed
with JHK$_s$, which can be interpreted as the consequence of an
inhomogeneous interstellar absorption, the UBV and
JHK$_s$ photometry gives an opportunity to estimate the reddening of the
cluster and to test at the same
time our temperature scale. By inverting the empirical relations $\Teff =
f(\mbox{colour}, \FeH)$
calibrated by Alonso et al. (1996) for dwarfs and by Alonso et al. (1999)
for giants, the colour index B$-$V
and V$-$K corresponding to the TGMET effective temperatures were computed
and compared to the observed ones,
adopting K$_s$ for K. A systematic difference between them
can be interpreted either in terms of reddening or as an error in the
temperature scale. Giants have
been tested first because their TGMET temperature scale is more reliable
than for fast rotators.
The mean observed colours B$-$V and V$-$K for the cluster's giants are
respectively 1.04 and 2.54, for a mean
effective temperature of 5000~K. Such a temperature at $\FeH = -0.07$
corresponds to B$-$V=0.88 and V$-$K=2.12
in the Alonso et al.'s temperature scale. The corresponding excesses
$E_{B-V}=0.16$ and $E_{V-K}=0.42$ lead to a
ratio $E_{V-K}/E_{B-V}=2.62$ which agrees, within the error bars, with the
value of 2.7 reported by
Rieke \& Lebofsky (1985) and Cardelli et al. (1989) for the interstellar
extinction.
For the hot stars the ratio was slightly different, possibly indicating an
error in the temperature scale. The
mean observed V$-$K of the dwarfs (1.00), corrected by $E_{V-K}=0.42$ leads to
$\Teff$=7500~K with Alonso et al.'s
relations while the mean temperature estimated by TGMET is 7300 K. An
offset of 200 K in $\Teff$ is still
consistent with an age of 1 Gyr.

The individual distance moduli of the cluster members in Tab. 2 yield a mean
distance modulus of $11.56 \pm 0.10$
for the cluster. By correcting for interstellar absorption according to a
mean reddening of $E_{B-V}=0.16$ and
$A_\mathrm{V}/E_{B-V}=3.09$ (Rieke \& Lebofsky 1985), we determine the
distance of NGC 2355 as $1650^{+80}_{-70}$ pc.
The corresponding height above the galactic plane is 340 pc.  The
dereddened distance modulus $(V-\Mv)_0=11.06$ is consistent with the one
derived by Ann et al. (1999), $(m-M)_0=11.4$, by isochrone and ZAMS 
fittings on colour-magnitude diagrams
whereas they find a lower metallicity ($\FeH = -0.32$)  and a higher reddening 
$(E_{B-V}=0.25)$. On the contrary, we are in disagreement with Kaluzny \& Mazur 
(1991) for the distance modulus ($(m-M)_0=12.1$) but in better agreement for the metallicity 
and reddening ($\FeH=+0.13, E_{B-V}=0.12$). Isochrone and ZAMS fitting is well  
adapted for dense clusters with high quality multicolour photometry because
of the three parameters to be deduced simultaneously : age, metallicity and 
reddening. In the
case of NGC 2355, where the photometry is dispersed, this method can lead to 
a wide range of parameters as can be seen from the compared studies of 
Ann et al. (1999) and Kaluzny \& Mazur (1991). Spectroscopy concerns less stars 
but better constrains the parameters.
In Sect. 2, the angular radius of the cluster's central body was estimated
to be $7'$, while that of
its halo was estimated to be $15'$. At the distance of 1.65~kpc this
corresponds
to linear radii of 3.3~pc and 7.2~pc respectively.
The radius of the central body, 3.3 pc, is typical of the old open clusters linear radii which
are distributed in a small range with a median at 2.65 pc, and an upper quartile at
3.45 pc (Janes \& Phelps 1994).

Fig. 8 presents the dereddened colour magnitude diagram of the cluster in
(V$-$K,V), including
the members which have been confirmed by their radial velocity, and the
candidates within $7'$ of the
cluster's center having a probability higher than 90\% to be member on the
basis of proper motion.
For comparison the 1 Gyr isochrone has been transformed into observable
quantities and overlayed.
The bluest star, GSC~501264, in the prolongation of the main sequence, is a
typical  blue straggler candidate.
According to its colour $(J-K)_0 = 0.03$, Alonso et al.'s calibration gives
an effective temperature of 8300~K.
The colour index $(J-H) = 0.09$, for which the reddening is unknown, gives
a consistent temperature but
$(V-K)_0=0.19$ is unfortunately outside the limits of the calibration. It was not
possible to estimate spectroscopically such
a high temperature with TGMET because of the limit of the library to
7500 K.
The effective temperature of blue stragglers seems to correspond to a mass
which is higher
than that of the turnoff stars,  consequently unconsistent with the age of
the cluster. This phenomenon
was already mentioned in the case of the red giant KM~1 in Sect. 4.

\begin{figure}[h]
\resizebox{8.5cm}{!}{\includegraphics{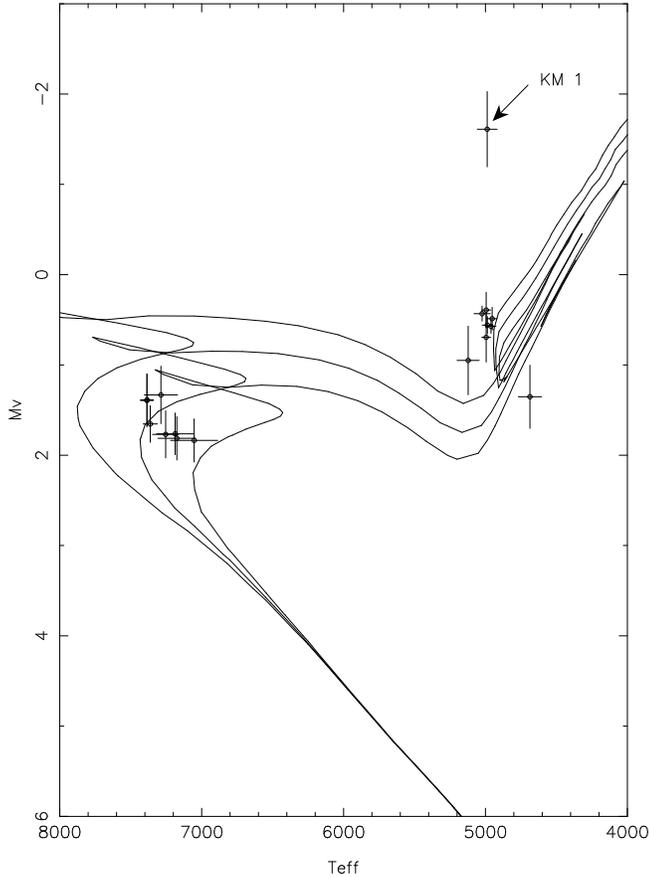}}
\caption{HR diagram of NGC 2355 from the TGMET parameters. Superposed are
the solar metallicity isochrones of
Girardi et al. (2000) corresponding to log(age yr)= 8.9, 9.0, 9.1.}
\label{}
\end{figure}

\begin{figure}[h]
\resizebox{8.5cm}{!}{\includegraphics{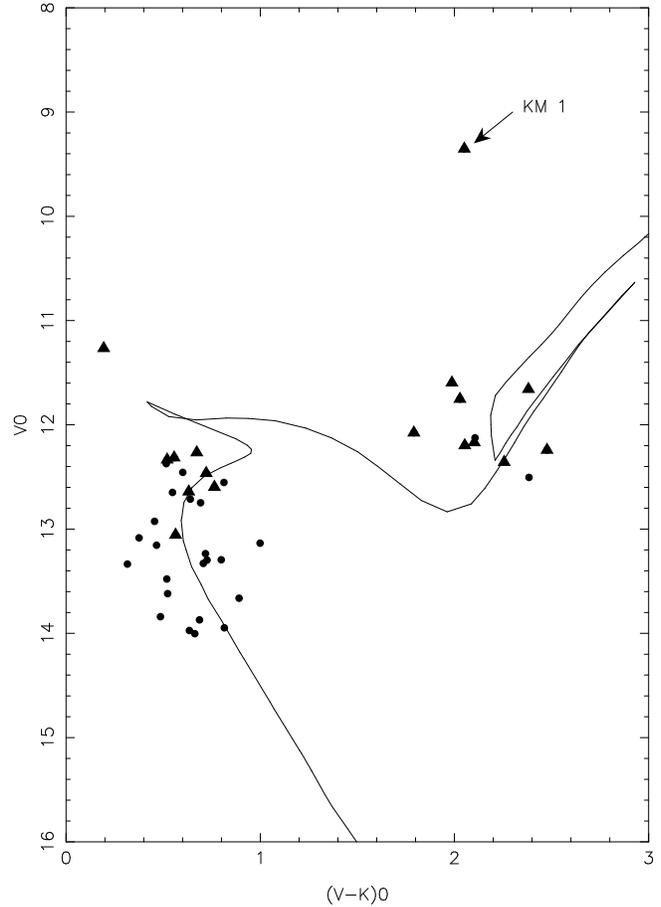}}
\caption{Dereddened colour-magnitude diagram of NGC 2355. Triangles
correspond to radial velocity members,
circles to proper motion members. The solid line corresponds to the solar
metallicity 1 Gyr isochrone of
Girardi et al. (2000).}
\label{}
\end{figure}

\subsection{Space velocity and galactic orbit}
Combining our results for the absolute proper motion, radial velocity and
distance of the cluster,
we determine its heliocentric space motion as $(U,V,W) = (-33.5\pm 1.8,
-18.8\pm 3.6,
-11.2\pm 3.9)\kms$
\footnote{To be clear, U,V,W are vector components with respect to a
right-handed triad pointing to galactic center, direction of rotation
and northern galactic pole}.
The uncertainties in the components of the space motion result from
the combination of all estimated observational errors which were given
in the previous sections. However, due to the relatively large distance of the
cluster from the Sun, the error budget is dominated by the uncertainty
in the cluster's proper motion.

In order to obtain the velocity of the cluster in the galactocentric frame,
we assume the motion of the Sun in the LSR as $(U,V,W)_\odot =
(9.7,5.2,6.7)\kms$,
following the recent result of Bienaym\'e (1999) which is supported by
similar results
of e.g. Dehnen \& Binney (1998).
Furthermore we adopt the current IAU standard values of $V_{LSR} = 220\kms$
for the local circular rotation velocity and $R_\odot = 8.5$ kpc for the
distance of the Sun
from the galactic center. The galactocentric position and velocity of the
cluster then is
$(x,y,z) = (-10.00, -0.64, +0.34)\kpc $ and $(U,V,W) = (-23.5, +206.2,
-4.2)\kms$.
Together with the galactic gravitational potential these vectors determine
the orbit of the cluster in the Galaxy.

We have integrated the equations of motion in the galactic model of Allen
\& Santillan (1991)
over the estimated cluster age of 1 Gyr. The resulting orbit is
characterized by radial oscillations
between distances from the galactic center of 8.9 and 10.1~kpc and vertical
oscillations with an
amplitude of 350 pc. The median of the distance $r$ from the galactic
center along the orbit
is 9.6~kpc and the median of the distance $|z|$ from the galactic plane is
0.24~kpc.
The cluster has made 3.7 revolutions around the galactic center and 21
crossings of the galactic disk
within its lifetime.
If one varies the measured space velocity of the cluster within the error bars
of the observations the parameters of the orbit undergo relatively
small changes. We find that the radial distances can differ by $\pm 3\%$
and the vertical distances by $\pm 7\%$ from the above given values for the
`mean orbit'.
Thus we can say with certainty that the cluster keeps well outside the
solar circle throughout
its revolution around the Galaxy.

We currently observe the cluster close to its maximum distance from the plane,
i.e. close to the point of reversal of the vertical oscillation.
This is consistent with the characteristics of the vertical motion because
the probability
to find the cluster near the maximum of $|z|$ (at a randomly chosen
instant) is about a factor
of three larger than the corresponding probability for a lower value of $|z|$.
The statistics of $|z|$ along the orbit is such that the cluster spends
only 9\% of its time in the
thin layer of the young disk population at $|z| \le 50\pc$ where close
encounters with very massive
molecular clouds could have occurred.
On the other hand, the orbit of NGC 2355 does not reach such extreme vertical
distances as a few other old open clusters which are found at $|z|$ up
to 2.4 kpc (Friel 1995).
Thus we recognize NGC 2355 as a fairly normal representative of the old open
cluster population.
The latter has a scale height of 375~pc (Janes \& Phelps 1994) as compared
to the scale height of 55~pc for the young population of open clusters.

\section{ Conclusion }
We have presented a detailed study of stars in the region of NGC 2355,
combining new astrometric
and spectroscopic data with recent photometric data from other sources.
Our main results can be summarised as follows :

- NGC 2355 is at 1.65~kpc of the Sun and 340~pc above the galactic plane in
the direction of the
  anticenter, with a reddening of $E_{B-V}=0.16$ and $E_{V-K}=0.42$.

- Its metallicity is $\FeH = -0.07 \pm 0.11$ and its age is 1 Gyr.

- NGC 2355 has a core radius of about 0.7~pc, a central component with a
radius of 3.3~pc and a
  halo out to 7.2~pc from the cluster center.

- The turnoff stars of NGC 2355 are fast rotators, with a mean projected
rotational velocity
  of $100\kms$ and a mean $\Teff$ of 7500 K.

- The giant clump is well defined at $\Teff = 5000 K$, $\Mv = 0.51$.

- Two stragglers have been identified in the cluster: a blue one, and a
giant which has an unusual
  position in the HR diagram, 2.3 mag brighter than the giant clump.

- NGC 2355 has a galactocentric space velocity vector $(U,V,W) = (-23.5,
+206.2, -4.2)\kms$ and an
  orbit which keeps it beyond the solar circle and with only brief passages
  through the galactic plane.

 As a by-product of the study of the cluster, we found a moving
pair of field giants with a radial
velocity of $50\kms$.

\begin{acknowledgements}
We thank J.Guibert and the MAMA team at Paris Observatory for their support by
scanning plates, R. LePoole from Leiden Observatory for lending the POSS-I
glass copies, the Astronomical Institute M\"unster for scanning time on the
PDS machine,
and all colleagues of Bordeaux Observatory who have taken part in the
meridian circle observations.
We also thank C. Catala who provided his observations to increase the TGMET
library,
J.-F. Donati who kindly made his deconvolution software available for us, J.-L.
Halbwachs and S. Piquard who
provided some intermediate measurements of TYCHO, and A. Alonso who
provided his calibrations before
publication.
We are also grateful to R. Cayrel for his comments and suggestions. We have
made use in this research
of the SIMBAD and VIZIER databases, operated at CDS, Strasbourg, France.
M.O. gratefully acknowledges financial support by a Marie Curie research
grant from the
European Community during this work.

\end{acknowledgements}

\end{document}